\documentstyle[prl,aps,multicol]{revtex}
\begin{document}
\title{Feynman Integrals with Absorbing Boundaries}
\author{A. Marchewka\\
School of Physics and Astronomy \\
Beverly and Raymond Sackler Faculty of Exact Sciences\\
Tel Aviv University\\
Tel Aviv 69978, Israel. e-mail: marhavka@post.tau.ac.il\\
~\\
Z. Schuss\\
Department of Mathematics\\
Tel-Aviv University\\
Ramat-Aviv, Tel-Aviv 69978, Israel. e-mail: schuss@math.tau.ac.il}
\date{\today }
\maketitle 
~\\
We propose a formulation of an absorbing boundary for a quantum particle.
The formulation is based on a Feynman-type integral over trajectories that
are confined to the non-absorbing region. Trajectories that reach the
absorbing wall are discounted from the population of the surviving
trajectories with a certain weighting factor. Under the assumption that
absorbed trajectories do not interfere with the surviving trajectories, we
obtain a time dependent absorption law. Two examples are worked out.\\


\begin{multicols}{2}
The purpose of this letter is to propose a Feynman-type integral to describe
absorption of particles in a surface bounding a domain. The need for such
description arises for example in scattering theory, in the description of a
photographic plate, in the double slit experiment, in neutron optics, and so
on. The optical model \cite{Hodgson,Sears} is often used to describe
absorption in quantum systems. This model is based on analogy with
electro-magnetic wave theory. It is not obvious that the methods of
describing absorption in Maxwell's equations carry over to quantum mechanics
because the wave function of a particle does not interact with the medium
the way an electro-magnetic wave does. In particular, in classical quantum
theory, unlike in electromagnetic theory, the wave function does not
transfer energy to the medium.

The main tenet of our Feynman-type integral description of absorption is
that trajectories that propagate into the absorbing surface for the first
time are considered to be instantaneously absorbed and are therefore
terminated at that surface. The population of the surviving trajectories is
therefore discounted by the probability of the absorbed trajectories at each
time step.

The process of discounting can be explained as follows. In general, if the
trajectories are partitioned into two subsets, the part of the wave function
obtained from the Feynman integral over one cannot be used to calculate the
probability of this subset, due to interference between the wave functions
of the two subsets. However, in the physical situation under consideration,
such a calculation may be justified. Our procedure, in effect, assumes a
partition of all the possible trajectories at any given time interval $[t,t+$
$\Delta t]$ into two classes. One is a class of {\em bounded trajectories}
that have not reached the surface by time $t+\Delta t$ and remain in the
domain, and the other is a class of trajectories that hit the surface for
the first time in the interval $[t,t+$ $\Delta t]$. We assume that the part
of the wave function obtained from the Feynman integral over trajectories
that hit the surface in this time interval no longer interferes with the
part of the wave function obtained from the Feynman integral over the class
of bounded trajectories in a significant way. That is, the interference is
terminated at this point so that the general population of trajectories can
be discounted by the probability of the terminated trajectories. This
assumption makes it possible to calculate separately the probability of the
absorbed trajectories in the time interval $[t,t+$ $\Delta t]$.

The discounting process constitutes a coarse-graining procedure for a large
quantum system, describing the absorber. The assumptions we make can be
viewed as the mathematical expression of quantum irreversibility of
absorption, because the absence of interference separates the two classes of
trajectories for all times. Thus the trajectories that stop to interfere can
be discounted from the population of trajecotires inside the domain. Our
model is only one aspect of irreversible processes.

Under the above assumptions the discounting procedure leads to a Feynman-Kac
integral with a killing measure \cite{book}, which in turn leads to a
Schr\"{o}dinger equation with zero boundary conditions on the absorbing
surface and complex valued energy which depends on the wave function of the
class of bounded trajectories at each time $t$.

We obtain a decay law that is not an exponential rate, in general. It
reduces to a rate only for a single energy level initial condition. In this
case, the wave function is a solution of the Schr\H{o}dinger equation with
infinite walls at the absorbing boundaries and constant complex potential
that depends on the initial energy. If, however, there are two initial
energy levels, the decay law depends on both energy levels and contains an
oscillatory term with beat frequency. Similar beats in the decay law occur
if the initial wave function contains any number, finite or infinite, of
energy levels.

We consider another example, of a Gaussian packet of free particles
traveling with a given mean velocity toward an absorbing wall. We find that
the packet is partially reflected and partially absorbed and calculate the
reflection coefficient. The reflected packet results from trajectories that
never reached the wall. This behavior is mainly due to classically forbidden
trajectories. In contrast to scattering, this is not the same reflection as
that in a finite or infinite barrier, because reflection in a finite barrier
depends on the shape of the barrier and reflection in an infinite barrier is
totally elastic, whereas the reflection we obtain is discounted by a finite
constant factor.

Consider the class $\sigma _{a,b}$ of continuously differentiable functions $%
x(\tau )$ for $0\leq \tau <\infty $ such that $a\leq x(\tau )\leq b$ for all 
$0\leq \tau <\infty $ and such that $x(0)=x_I,\ x(t)=x$. The class $\sigma
_{a,b}$ consists of bounded trajectories that begin at $x_I$ and end at $x$.
We define the Feynman integral over the class $\sigma _{a,b}$ by 
\begin{eqnarray}
&&K(x,t)=\int_{\sigma _{a,b}}\exp \left\{ \frac i\hbar S\left[ x(\cdot
),t\right] \right\} {\cal D}x(\cdot )  \label{Flab} \\
&&\equiv \lim_{N\rightarrow \infty }\alpha ^N\int_a^b\dots \int_a^b\exp
\left\{ \frac i\hbar S(x_0,\dots ,x_N,t)\right\} \prod_{j=1}^{N-1}\,dx_j, 
\nonumber
\end{eqnarray}
where 
\[
\alpha =\left\{ \frac m{2\pi i\hbar \Delta t}\right\} ^{1/2}. 
\]

Next, following the method of \cite{Keller}, we show that $K(x,t)$ satisfies
Schr\"{o}dinger's equation and determine the boundary conditions at the
endpoints of the interval $[a,b]$. We begin with a derivation of a recursion
relation that defines $K(x,t)$. We set 
\begin{eqnarray*}
&&K_N(x_N,t)\equiv \\
&&\alpha ^N\int_a^b\dots \int_a^b\exp \left\{ \frac i\hbar S(x_0,\dots
,x_N,t)\right\} \prod_{j=1}^{N-1}\,dx_j,
\end{eqnarray*}
then, by definition, $K(x,t)=\lim_{N\rightarrow \infty }K_N(x,t)$. We have
therefore the recursion relation 
\begin{eqnarray}
K_N(x,t)&=&\alpha \int_a^b\exp \left\{ \frac i\hbar \left[ \frac{%
m(x-x_{N-1})^2}{2\Delta t}-V(x)\Delta t\right] \right\}  \nonumber \\
&&  \nonumber \\
&\times&K_{N-1}(x_{N-1},t_{N-1})\,dx_{N-1}.  \label{recursion}
\end{eqnarray}
The following derivation is formal, a strict derivation can be constructed
along the lines of \cite{Keller}. We expand the function $%
K_{N-1}(x_{N-1},t_{N-1})$ in (\ref{recursion}) in Taylor's series about $x$
to obtain 
\begin{eqnarray}
\ &&K_N(x,t)=  \label{Taylor} \\
&&\ \alpha e^{-iV(x)\Delta t/\hbar }\int_a^b\exp \left\{ \frac{im}{2\hbar
\Delta t}(x-x_{N-1})^2\right\} \times  \nonumber \\
&&\ \left[ K_{N-1}(x,t_{N-1})-(x-x_{N-1})\frac{\partial K_{N-1}(x,t_{N-1})}{%
\partial x}+\right.  \nonumber \\
&&\ \frac 12(x-x_{N-1})^2\frac{\partial ^2K_{N-1}(x,t_{N-1})}{\partial x^2}+
\nonumber \\
&&\left. O\left( \left( x-x_{N-1}\right) ^3\right) \right] \,dx_{N-1}. 
\nonumber
\end{eqnarray}

We evaluate the integrals in eq.(\ref{Taylor}) separately for $x$ inside the
interval $[a,b]$ and on its boundaries. This leads to the Schr\"{o}dinger
equation

\begin{equation}
{i\hbar }\frac{\partial K(x,t)}{\partial t}=-\frac{\hbar ^2}{2m}\,\frac{%
\partial ^2K(x,t)}{\partial x^2}+V(x)K(x,t)\quad  \label{SE}
\end{equation}
for $a<x<b$ with boundary and initial conditions 
\begin{eqnarray}
K(a,t) &=&K(b,t)=0\quad \quad \mbox{for $t>0$}  \label{BC} \\
&&\mbox{}  \nonumber \\
K(x,0) &=&\delta (x-x_I)\quad \quad \mbox{for $a<x<b$}.  \label{ic}
\end{eqnarray}
Obviously, eqs.(\ref{SE})-(\ref{ic}) are identical to those of a particle
bounded by infinite potential walls.

The same result was obtained in \cite{Kleinert1} for the case $V(x)=0$ by a
different method. Our method of calculation is essential for calculating the
Feynman integral with absorbing boundaries.

First, we calculate the discretized Feynman integral to survive (not to be
absorbed) the time interval $\left[ 0,\Delta t\right] $ and to find a
trajectory in time $\Delta t$ at a point $x$ in the interval $\left[
a,b\right] $. According to the above assumptions, the discretized Feynman
integral for trajectories initially inside the interval $\left[ a,b\right] $
that propagate to the endpoint $a$ for the first time in the time interval $%
\left[ 0,\Delta t\right] $ is 
\[
\psi _1(a,\Delta t)=\alpha \int_a^b\Psi _0(x_0)\exp \left\{ \frac i\hbar
S(x_0,a,\Delta t)\right\} \,dx_0. 
\]
Therefore, the probability density of finding a trajectory at the point $a$
in the time interval $\left[ 0,\Delta t\right] $ is 
\[
\left| \psi _1(a,\Delta t)\right| ^2, 
\]
and there is an analogous expression for the probability density of finding
a trajectory at the point $b$ in the time interval $\left[ 0,\Delta t\right]
.$ It follows that the probability of a trajectory to be absorbed in the
time interval $\left[ 0,\Delta t\right] $ is 
\[
P_1(\Delta t)=\lambda _a\left| \psi _1(a,\Delta t)\right| ^2\,+\lambda
_b\left| \psi _1(b,\Delta t)\right| ^2\,, 
\]
where $\lambda _a$ and $\lambda _b$ are characteristic lengths (see
discussion at the end of the letter). Thus the discretized Feynman integral
to survive the time interval $\left[ 0,\Delta t\right] $ and find a
trajectory in time $\Delta t$ at a point $x$ in the interval $\left[
a,b\right] $ is 
\begin{eqnarray}
&&\Psi _1(x,\Delta t)=\sqrt{1-P_1(\Delta t)}\alpha \int_a^b\Psi _0(x_0)\times
\nonumber \\
&&\exp \left\{ \frac i\hbar S(x_0,x,\Delta t)\right\} \,dx_0=  \nonumber \\
&&\sqrt{1-P_1(\Delta t)}\,K_1(x,\Delta t).  \label{psi1xt}
\end{eqnarray}

Next, we calculate the discretized Feynman integral to survive the time
interval $\left[ \Delta t,2\Delta t\right] $ and find a trajectory in time $%
2\Delta t$ at a point $x$ in the interval $\left[ a,b\right] $. According to
eq.(\ref{psi1xt}), given that a trajectory survived to time $\Delta t$, its
discretized wave function is $K_1(x,\Delta t)$ so that the discretized
Feynman integral to propagate to the point $a$ is 
\[
\psi _2\left( a,2\Delta t\right) =\alpha \int_a^bK_1(x,\Delta t)\exp \left\{ 
\frac i\hbar S(x,a,\Delta t)\right\} \,dx. 
\]

Proceeding this way, we find that the discretized Feynman integral to
survive the time interval $\left[ 0,N\Delta t\right] $ and find a trajectory
in time $N\Delta t$ at a point $x$ in the interval $\left[ a,b\right] $ is 
\begin{equation}
\Psi _N(x,N\Delta t)=\sqrt{\prod_{j=1}^{N-1}\left( 1-P_j(j\Delta t)\right) }%
\,K_N(x,t),  \label{psint}
\end{equation}
where 
\begin{equation}
P_j(j\Delta t)=\lambda _a\left| \psi _j(a,j\Delta t)\right| ^2\,+\lambda
_b\left| \psi _j(b,\Delta t)\right| ^2\,.  \label{pab}
\end{equation}

It remains to calculate the survival probability 
\begin{equation}
1-P(t)=\lim_{N\rightarrow \infty }\prod_{j=1}^{N-1}\left( 1-P_j(j\Delta
t)\right) .  \label{sp}
\end{equation}
It can be shown \cite{MS} that the probability $P_j(j\Delta t)$ is given by 
\begin{eqnarray*}
P_j(j\Delta t) &=&\frac{\hbar \Delta t}{2\pi m}\left[ \lambda _a\left| \frac %
\partial {\partial x}K_{j-1}\left( a,(j-1)\Delta t\right) \right| ^2+\right.
\\
&&\left. \lambda _b\left| \frac \partial {\partial x}K_{j-1}\left(
b,(j-1)\Delta t\right) \right| ^2+o(1)\right] ,
\end{eqnarray*}
so that eq.(\ref{sp}) gives 
\begin{eqnarray}
&&1-P(t)=  \label{1-pt} \\
&&\exp \left\{ -\frac \hbar {\pi m}\int_0^t\left[ \lambda _a\left| \frac %
\partial {\partial x}K\left( a,t\right) \right| ^2\,+\right. \right. 
\nonumber \\
&&\left. \left. \lambda _b\left| \frac \partial {\partial x}K\left(
b,t\right) \right| ^2\right] \,dt.\right\}
\end{eqnarray}
Now, it follows from eqs.(\ref{psint}) that the wave function of the
surviving trajectories at time $t$ is given by 
\begin{equation}
\Psi (x,t)=\sqrt{1-P(t)}K(x,t),  \label{WFN}
\end{equation}
and $1-P(t)$ is given by (\ref{1-pt}).\newline

\noindent
{\bf Examples}\newline

First, we consider a particle with two absorbing walls at $x=\pm a$ and zero
potential. We assume that $\lambda _{-a}=\lambda _a$. The wave function is
given by 
\[
K(x,t)=\sum_{n=1}^\infty A_n\exp \left\{ -\frac{-i\hbar n^2\pi ^2}{2ma^2}%
t\right\} \sin \frac{n\pi }ax 
\]
so that 
\begin{eqnarray*}
&&\int_0^t\left| \frac \partial {\partial x}K\left( \pm a,t\right) \right|
^2\,\,dt=\sum_{n=1}^\infty \sum_{k\neq n}^\infty \frac{A_k\bar{A}_n}{k^2-n^2}%
\frac{2knm}{i\hbar a^2}\left( -1\right) ^{k+n}\times \\
&&\left[ 1-\exp \left\{ -\frac{i\hbar \left( k^2-n^2\right) \pi ^2}{2ma^2}%
t\right\} \right] +\sum_{n=1}^\infty \left| A_n\right| ^2\frac{n^2\pi ^2}{a^2%
}t.
\end{eqnarray*}

For a particle with a single energy level the wave function decays at an
exponential rate proportional to the energy. However, if there are more than
just one level, the exponent contains beats. For example, for a two level
system with real coefficients, we obtain 
\begin{eqnarray*}
1-P(t) &=&\exp \left\{ -\frac{\lambda _a\hbar }{\pi m}\left[ \frac{\pi ^2}{%
a^2}\left( A_k^2k^2+A_n^2n^2\right) t-\right. \right. \\
&&\left. \left. \frac{4mA_kA_n}{\hbar \left( k^2-n^2\right) }\sin \frac{%
\hbar \left( k^2-n^2\right) \pi ^2}{2ma^2}t\right] \right\} .
\end{eqnarray*}
The strongest beats occur for $k=2,\,n=1$ with frequency $\omega _{1,2}=%
\frac{3\hbar \pi ^2}{2ma^2}$. Setting $A_1=A_2=\sqrt{1/2}$ and introducing
the dimensionless time $\tau =\frac{\lambda _a\hbar \pi }{ma^2}t,$ we find
that 
\begin{equation}
1-P(t)=\exp \left\{ -\frac 52\tau +\frac 2{3\pi }\sin \frac{3\pi }2\tau
\right\} .  \label{35}
\end{equation}

Next, we consider a Gaussian-like wave packet of free particles traveling
toward an absorbing wall at $x=0$ with positive mean velocity $k_0$. That
is, in order to maintain the zero boundary condition on the wall the initial
wave function is the difference between two antisymmetric Gaussians relative
to the absorbing wall. It follows that 
\begin{eqnarray}
&&\left| \frac \partial {\partial x}K(0,t)\right| ^2=\frac a{16\pi ^2}\frac{%
\frac{a^4}{16}k_0^2+x_0^2}{\left( \frac{a^4}{16}+\frac{t^2}{4m^2}\right) ^{%
\frac 32}}\times  \label{abk} \\
&&\exp \left\{ \frac{\frac{a^2}4\left( \frac{a^2}2k_0^2+x_0^2\right) -\frac{%
a^2}{4m}x_0k_0t}{\frac{a^4}{16}+\frac{t^2}{4m^2}}-\frac{a^2}2k_0^2\right\} ,
\nonumber
\end{eqnarray}
hence 
\begin{equation}
\int_0^\infty |\Psi ^{\prime }(0,t)|^2\,dt<\infty .  \label{FINT}
\end{equation}
Thus 
\[
R=\lim_{t\rightarrow \infty }\left[ 1-P(t)\right] >0, 
\]
that is, the wave packet is only partially absorbed. This means that the
``reflected'' wave consists of trajectories that turned around before
propagating into the absorbing wall where absorption occurs. The discount of
the wave function occurs when the packet is at the wall, as can be seen from
eqs. (\ref{abk}) and (\ref{FINT}). Thus $R$ plays the role of a {\em %
reflection coefficient. }This is neither the usual reflection coefficient
for a finite potential barrier nor that for an infinite barrier.

The two examples can be combined into a simple experimental setup of a
cavity with absorbing walls and an absorbing detector at one end. A particle
travelling along the axis of the cavity fits the first example in the the
transverse direction and the second example in the direction of the cavity
axis. Thus the decay law is the product of the two decay laws described
above. Further examples and applicatins are discussed in \cite{MS}.\newline

\noindent {\bf Discussion}\newline

Absorption in a surface is different than absorption in the bulk across the
surface in that Feynman trajectories do not propagate across the surface in
the former but do in the latter case. This letter is concerned with
absorption in a surface. Absorption in the bulk requires a separate theory.
The basic assumption in our model is that Feynman trajectories that
propagate into the surface are instantaneously absorbed and the probability
of the remaining trajectories is discounted by the probability of the
absorbed trajectories at each time step. Thus the instantaneous discount
factor is proportional to the probability density at the surface at each
time step. The proportionality constant, denoted $\lambda $, is a
characteristic length, in analogy with the scattering length, the mean free
path \cite{Sears}, or a typical Compton wavelength. It serves as a fudge
parameter in this theory and is expected to be a measurable quantity. It may
depend on the energy of the particles, on the temperature of the absorbing
medium, and so on.

Our derivation does not start with a Hamiltonian, but rather with an action
of bounded trajectories. The resulting decaying wave function corresponds to
a classical quantum system with a Hamiltonian whose potential is complex
valued and time dependent. This potential depends on the initial energies of
the system\cite{foot}. Thus, in our formalism, the trajectories are given an actual
physical interpretation as the possible trajectories of a quantum particle.
This is analogous to the trajectory approach to diffusion in probability
theory \cite{Freidlin,Schulman,Kleinert}.

Quantum mechanics without absorption is recovered from our formalism when
the absorbing boundaries are moved to infinity. In higher dimensions,
quantum mechanics without absorption can be recovered from our formalism by
putting absorbing regions with variable density in a half space, say. As the
density increases, the boundary of the half space becomes a totally
absorbing wall and as the density decreases to zero, quantum mechanics is
recovered.

The examples demonstrate the expected phenomenon that particles that reach
the absorbing boundary are partially reflected and partially absorbed. In
either case the decay pattern of the wave function seems to be new.\newline

\noindent {\bf Acknowledgment}: The authors thank Yakir Aharonov for useful
discussions.

\end{multicols}
\end{document}